\documentclass[aps,reprint,pra]{revtex4-1}
\usepackage{graphicx}
\usepackage{dcolumn}
\usepackage{bm}
\usepackage{amsmath}
\usepackage{latexsym}
\numberwithin{equation}{section}

\begin{document}

\def\calD{{\cal D}}
\def\bfx{{\bf x}}
\def\bfr{{\bf r}}
\def\bfR{{\bf R}}

\title{Liquid-state polaron theory of the hydrated electron revisited}
\author{James P. Donley}
\email{jdonley@valence4.com}
\affiliation{Valence4 Technologies, Arlington, VA 22202}
\author{David R. Heine}
\affiliation{Corning, Inc., Corning, NY 14830}
\author{Caleb A. Tormey}
\affiliation{Dept. of Chemistry and Geochemistry, Colorado School of Mines, Golden, CO 80401}
\author{David T. Wu}
\affiliation{Dept. of Chemistry and Geochemistry, Colorado School of Mines, Golden, CO 80401}

\date{\today}

\begin{abstract}
The quantum path integral/classical liquid-state theory of Chandler and co-workers, created to describe an excess electron in solvent, is re-examined for the
hydrated electron.
The portion that models electron-water density correlations is replaced by two
equations: the
range optimized random phase approximation (RO-RPA), and the DRL approximation
to the ``two-chain'' equation,
both shown previously to describe accurately the static structure and
thermodynamics of strongly charged polyelectrolyte solutions. The static
equilibrium properties of the hydrated electron are analyzed using
five different electron-water pseudopotentials. The theory is then compared with
data from mixed quantum/classical Monte Carlo and molecular dynamics
simulations using these same pseudopotentials.
It is found that the predictions of the RO-RPA and DRL-based polaron theories
are similar and improve upon previous theory, with values for almost all
properties analyzed in reasonable quantitative agreement with the available
simulation data. Also, it is found using the Larsen, Glover and Schwartz pseudopotential that the theories give values for the solvation free energy that are
at least three times larger than that from experiment.
\end{abstract} 


\maketitle

\section{\label{sec:intro}Introduction}
The solvated electron, the smallest anion, is a highly reactive species, with
a lifetime $\sim 10^{-3}$ seconds in pure water at room temperature.\cite{mozumder99} It is an intermediate in many water-based chemical reactions.\cite{buxton88}

A large part of our understanding of the hydrated electron has come from theory and simulation.\cite{feng80,sprik86,schnitker87b,barnett88b,turi02,jacobson10,laria91,turi12}
Most of the modern efforts have modeled the electron-water and water-water interactions
through pseudopotentials.
Results of this approach and others have led to a view that the hydrated electron is a localized, self-trapped object. Recently, however, this view  has been
challenged,\cite{larsen10} and the subject is currently
controversial.\cite{turi11,jacobson11,larsen11,herbert11,turi12,uhlig12,casey13}

Our primary interest here though is with simpler theoretical approaches, the
aim being ultimately to go beyond the basic system of a single electron in
infinite water, to study bipolarons for example.\cite{laria91}
The question then is whether, given a particular pseudopotential, the theory
can give accurate results in comparison with more complex methods such as mixed
quantum/classical molecular dynamics simulation.

One successful theory to date has been the RISM-polaron one of Chandler and co-workers.\cite{chandler84,nichols84,laria91} In their work
on the hydrated electron, Laria, Wu and Chandler (LWC) found that the
theory gave reasonable predictions in comparison with simulation for quantities
such as the excess chemical potential and electron polymer size. However, its
modeling of electron-water density correlations was not as good as one would
 expect, local structure usually being a strong point
of liquid-state theory.\cite{hansen86,schweizer97,curro99,heine05}
LWC concluded that the equation developed by them for the electron-water radial
distribution function was the primary cause. Also, in some cases,
the theory predicted a ``super-trapped'' electron polymer, but there has been
no subsequent supporting evidence for its existence.

Since the work of LWC, two liquid-state theories, the range optimized
random phase approximation (RO-RPA)\cite{donley04a} and the approximation of
Donley, Rajasekaran and Liu (DRL)\cite{donley98} to the ``two-chain'' equation,
have been developed. Both theories have been shown to give quantitatively
accurate descriptions of the structure and thermodynamics of strongly charged
polyelectrolyte solutions.\cite{donley04a,donley06} In particular, it was
found that site-level correlations included in these theories, but not
captured well by the LWC liquid-state element, were important. 

One purpose of the present work then is to revisit the RISM-polaron theory
of the hydrated electron, but instead use the RO-RPA and DRL equations for the
electron-water radial distribution functions. Employing a few
pseudopotentials should be sufficient to determine the accuracy of the theory
and which elements need improvement. However, there are important qualitative
differences
among the candidate pseudopotentials. So using
theory to explore other phenomena is suspect until this controversy is settled.
A second purpose then is to compare with as many 
pseudopotentials involved in the current discussion as possible.
As will be shown below, the theory can shed
light on some questions that have not been convincingly answered as yet.

The remainder of this paper is organized as follows. In Sec. \ref{sec:theory}
the basic elements of the RISM-polaron theory, and
then the RO-RPA and DRL theories, are described briefly. In
 Sec. \ref{sec:other_ingredients} various other ingredients needed by the theory
are discussed, including the five pseudopotentials used in the analysis.
In Sec. \ref{sec:numerics} the numerical procedure to solve the theory is
described, and in Sec. \ref{sec:results} results are presented. Finally in
Sec. \ref{sec:summary} the work is summarized
and directions for further work are discussed.

\section{\label{sec:theory}Theory}
\subsection{\label{subsec:rism_polaron_theory}Review of RISM-polaron theory}
In this section, the RISM-polaron theory of Chandler and co-workers
\cite{chandler84,nichols84,laria91} is reviewed briefly, with the main
approximations of the theory highlighted. The notation is the same as for
LWC where relevant, except in a few instances.

An important thermodynamic quantity for a solvated electron 
is the solvation free energy or excess chemical potential, $\Delta \mu$, which
is the work needed
to add a free electron to the solvent. In terms of the electron-solvent
equilibrium free energy, $F$,
\begin{equation}
\Delta\mu = F - F_{free},
\label{eq:deltamu}
\end{equation}
where $F_{free}$ is the free energy of the electron and solvent system when the electron is free and far away. In the canonical ensemble,
\begin{equation}
\beta F = -{\rm ln}\ Z,
\label{eq:free_energy}
\end{equation}
where $\beta$ is the inverse of $k_BT$ with $k_B$ being Boltzmann's
constant and $T$ the absolute temperature. The partition function is
\begin{equation}
Z ={\rm Tr}[\rho_S],
\label{eq:partition_function}
\end{equation}
with the statistical density matrix element:\footnote{The
usual density matrix element is $\rho_S/Z$.}
\begin{equation}
\rho_S(\bfx,\{\bfR\};\bfx^\prime,\{\bfR^\prime\}) = \langle \bfx,\{\bfR\}\vert e^{-\beta H} \vert\bfx^\prime,\{\bfR^\prime\}\rangle.
\label{eq:density_matrix}
\end{equation}
Here, $\bfx$ is the electron position and $\{\bfR\}$ is the set of solvent
coordinates, with $\bfR_{is}$ (denoted as $\bfr_i^{(s)}$ in LWC) being the
position of site $s$ on solvent molecule $i$.
Also, $H$ is the full electron and solvent Hamiltonian,
and $\vert\bfx,\{\bfR\}\rangle$ is an eigenket in the position basis of the
electron-solvent system.

It is convenient to evaluate $\rho_S$ using path integrals.
An especially appealing method is to integrate out the degrees of freedom of
the surrounding medium, leaving one 
that depends only on the electron position $\bfx$.\cite{feynman65}

Experiments involving electrons in water typically occur near room temperature.
At $T$=298 K, the electron thermal wavelength
 $\lambda_e = \sqrt{{\beta \hbar^2\over m}} \simeq 17.2$\AA, which is large. Here, $\hbar$ is Planck's constant
divided by $2\pi$ and $m$ is the electron mass. Thus, quantum effects must
be considered in determining its properties. On the other hand, the
thermal wavelength for water is around $0.1$\AA, which is much smaller than
the size of water and the range of the internal forces of the liquid. It 
seems reasonable then to treat the properties of water using classical
methods.

Treating the solvent classically enormously simplifies the integration
over their degrees of freedom. Chandler, Singh and Richardson (CSR) find, to
second order in the electron-solvent interactions,
that the effective diagonal statistical density matrix element for the
electron is:\cite{chandler84}
\begin{eqnarray}
\rho_S(\bfx,\bfx) = &&\ C_1\int\calD\{\bfR\}
\rho_S(\bfx,\{\bfR\};\bfx,\{\bfR\})\nonumber \\
\simeq &&\ C_2\int_\bfx^\bfx\calD[\bfr(\tau)] exp(S[\bfr(\tau)]),
\label{eq:rho_electron}
\end{eqnarray}
where $\int\calD\{\bfR\}$ denotes an integral
over the space of solvent positions $\{\bfR\}$, $C_1$ and $C_2$ are
constants, and
 $\int_\bfx^\bfx\calD[\bfr(\tau)]$ denotes an integral over all electron
paths $\bfr(\tau)$ in imaginary time $\tau$, $0\leq \tau\leq \beta\hbar$ that
start and end at position $\bfx$. The imaginary time action is
\begin{equation}
S=S_0 + S_I
\label{eq:action}
\end{equation}
where the electron kinetic energy term is
\begin{equation}
S_0[\bfr(\tau)] = -{m\over 2\hbar}\int_0^{\beta\hbar} d\tau \bigl\vert{\dot\bfr(\tau)}\bigr\vert^2,
\label{eq:S0}
\end{equation}
the dot denoting a time derivative, and the solvent-mediated electron-electron
interaction term is
\begin{eqnarray}
S_I&&[\bfr(\tau)] = -\rho\sum_s \beta{\hat u}_{es}(0) - \nonumber \\
&&\ {1\over 2(\beta\hbar)^2}\int_0^{\beta\hbar}d\tau\int_0^{\beta\hbar}d\tau^\prime v(\vert \bfr(\tau)-\bfr(\tau^\prime)\vert).
\label{eq:SI}
\end{eqnarray}
Here, $\rho = N/V$ is the average solvent molecular number density with
$N$ being the number of solvent molecules in the liquid volume $V$.
The term ${\hat u}_{es}(0)$ is the zero wavevector value of the Fourier
transform of $u_{es}(r)$, which is the (pseudo-)potential between the electron
and a site $s$ on the solvent molecule. For the water pseudopotentials used 
in this work, the index $s$ takes three values corresponding to the oxygen
and two hydrogen atoms.

The medium-induced potential has the familiar RPA form:
\begin{equation}
v(r) = - \beta^2\sum_{s,s^\prime}u_{es}*\chi_{ss^\prime}*u_{s^\prime e}(r),
\label{eq:mip}
\end{equation}
where the asterisks (*) denote convolutions and $r=\vert\bfr\vert$. Also,
 $\chi_{ss'}(r)$ is
the density-density correlation function between solvent sites $s$ and $s'$: 
\begin{equation}
\chi_{ss'}(r) = \bigl\langle ({\hat\rho}_s(\bfr)-\rho)
({\hat\rho}_{s'}(0)-\rho)\bigr\rangle,
\label{eq:chi}
\end{equation}
where the brackets denote an ensemble average, and the microscopic density
of solvent site $s$ is
\begin{equation}
{\hat\rho}_s(\bfr) = \sum_{i=1}^N \delta(\bfr - \bfR_{is}),
\label{eq:rhohat}
\end{equation}
with $\delta(\bfr)$ being the Dirac ``delta'' function at position $\bfr$.

The RPA, as a generalization of Debye-H{\" u}ckel theory, is 
strictly valid only for weak interactions, which would appear not to be the
case for the solvated electron. As such, CSR give arguments for extending the
validity of Eq.(\ref{eq:mip}) to much stronger interactions by
optimizing $S_I$ through the substitution of the bare potential
$-\beta u_{es}(r)$ with the direct correlation function $c_{es}(r)$, defined
by the RISM equation.\cite{chandler72} For this system, the RISM equation is
\begin{equation}
\rho h_{es}(r) = \sum_{s'}\omega_{e}*c_{es'}*\chi_{s's}(r),
\label{eq:RISM}
\end{equation}
where $h_{es}(r) = g_{es}(r) - 1$, with the electron-solvent radial
 distribution function,
\begin{equation}
g_{es}(r) = {1\over \beta\hbar}\int_0^{\beta\hbar}d\tau\ {1\over N}\sum_{i=1}^N\langle V\delta(\bfr-\bfr(\tau)+\bfR_{is})\rangle,
\label{eq:gofr_def}
\end{equation}
where the brackets denote a thermal average over the configurations of the
electron and solvent.
Note that this is a site-site function, rather than being defined relative
to the electron center-of-mass.

The function $\omega_e(r)$ in Eq.(\ref{eq:RISM}) describes site-site electron
self-correlations. It is defined as
\begin{equation}
\omega_{e}(r) = {1\over \beta\hbar}\int_0^{\beta\hbar}d\tau\ \omega_e(r,\tau),
\label{eq:omegaofr}
\end{equation}
where
\begin{equation}
\omega_e(\bfr,\tau-\tau^\prime) =
 \langle\delta(\bfr - \bfr(\tau) + \bfr(\tau^\prime))\rangle,
\label{eq:omegartau}
\end{equation}
the brackets denoting an average over all paths $\bfr(\tau)$ weighted
by the imaginary time action given by Eq.(\ref{eq:action}) above.

In this manner, 
\begin{equation}
v(r)\rightarrow -\sum_{ss'} c_{es}*\chi_{ss'}*c_{s'e}(r).
\label{eq:mip2}
\end{equation}
For polymer molecules, this hypernetted-chain (HNC)\cite{hansen86} form for the
 medium-induced potential was derived by
Melenkevitz, Schweizer and Curro\cite{melen93} using the density functional
theory of Chandler, McCoy and Singer (CMS)\cite{chandler86}. The
substitution of $-\beta{\hat u}_{es}(0)$ in Eq.(\ref{eq:SI}) with
${\hat c}_{es}(0)$ is an ansatz.

The approximate integrating out of the degrees of freedom of the solvent as
done by CSR is unusual in that these degrees of freedom are considered
``slow''\cite{chandler84} compared with the fluctuations of the electron.\footnote{In the original application of the path integral method to 
polarons by Feynman, the degrees of freedom of the surrounding crystal
were also slow. In that case though, the slow phonon modes were integrated out
exactly, they being modeled in the standard way as harmonic oscillators} Instead, the
fast degrees of freedom are usually integrated out, they then providing
a smooth background force and thermal bath for the slower
degrees of freedom of interest.\cite{langer71} At the least, the degrees of
freedom with similar relaxation timescales are integrated out, such as for
polymer melts and solutions.\cite{melen93}
For example, in many mixed quantum/classical Monte Carlo and
molecular dynamics (MD) treatments of the hydrated electron, the electron motion
is approximated as responding instantaneously to any change in the surrounding
solvent.\cite{barnett88b} This approximation is opposite to that done by CSR.
So it can be expected for the hydrated electron that, by truncating the
medium-induced potential at second order in the potential $u_{es}(r)$ as 
done above, some of the anisotropy of the 
electron-water correlations is sure to be lost.\cite{laria91}
Though to what degree will be shown partly in the results to follow.

To compute the free energy and electron self-correlation function, the
effective one-electron path integral, Eq.(\ref{eq:rho_electron}) can be
evaluated by Monte Carlo simulation.\cite{wuThesis,sumi04} However, a popular
alternative is to define an effective interaction action,\cite{feynman62}
\begin{equation}
S_{ref} = -{1\over 2}\sum_{n\neq 0}\gamma_n \vert \bfr_n\vert^2,
\label{eq:Sref}
\end{equation}
where the mode $\bfr_n$ is the Fourier transform of the 
electron imaginary time position $\bfr(\tau)$:
\begin{equation}
\bfr_n = {1\over \beta\hbar}\int_0^{\beta\hbar}d\tau\ \bfr(\tau)e^{i\Omega_n\tau},
\label{eq:rn}
\end{equation}
with frequencies $\Omega_n = 2\pi n/(\beta\hbar)$, and $\gamma_n$
is a mode dependent constant.

Then, the action $S = S_0 + S_{ref} + \Delta S$, with $\Delta S = S_I-S_{ref}$.
With $\Delta S=0$, the path integral can be solved exactly analytically. In
this way, the free energy is expanded to first order in $\Delta S$ yielding
an expression that can be minimized with respect to the
coefficients $\gamma_n$ to obtain an upper bound on the true free energy.
The resultant equation for the $\gamma_n$ is:\cite{chandler84}
\begin{eqnarray}
\gamma_n =&& -{1\over 6\pi^2}\int_0^\infty dk\ k^4 {\hat v}(k)\times\nonumber \\
&& {1\over \beta\hbar}\int_0^{\beta\hbar}d\tau\ {\hat\omega}_e(k,\tau)[1 - \cos(\Omega_n\tau)],
\label{eq:gamman}
\end{eqnarray}
where ${\hat v}(k)$ and ${\hat\omega}_e(k,\tau)$ are the Fourier transforms of the
medium-induced potential Eq.(\ref{eq:mip2}), and time dependent electron
self-correlation function Eq.(\ref{eq:omegartau}), respectively. The
latter function is computed self-consistently using the
effective action $S_0+S_{ref}$. Note that the direct
correlation function $c_{es}(r)$ also varies with changes in the electron 
self-correlations, yet terms due to that don't appear in Eq.(\ref{eq:gamman}),
even though they are large. The reason is that the variations of 
$c_{es}(r)$ in the
two terms of $S_I$, Eq.(\ref{eq:SI}) cancel. So, the ansatz of
$-\beta{\hat u}_{es}(0) \rightarrow {\hat c}_{es}(0)$ is an important one.

Feynman's path integral polaron formalism has been used in other areas at least since the work of Edwards on neutral polymers.\cite{doi86}
He and others have found that for real polymers, reference actions
 such as Eq.(\ref{eq:Sref}) need to be used with care as they can
 give erroneous results. For example,
des Cloizeaux showed that using Eq.(\ref{eq:Sref}), and evaluating its
coefficients via the free energy minimization scheme above gave incorrect
scaling for the size of a neutral polymer in solution.\cite{descloizeaux91}
The cause is thought to be the self-avoidance property of a real polymer.
So, while this problem appears not to be an issue for the electron,
this subtlety should be kept in mind.

Since $\omega_e(r)$ can be computed using the effective action
$S_0+S_{ref}$, the theory
is completed by specifying the electron-solvent potentials $u_{es}(r)$,
radial distribution functions $g_{es}(r)$, and the solvent density-density
correlation function $\chi_{ss^\prime}(r)$. The pseudopotentials used
will be discussed in Sec. \ref{subsec:pseudopotentials} below. The method of
 obtaining $\chi_{ss^\prime}(r)$ is very 
similar to LWC, and will be discussed in Sec. \ref{subsec:water_sim} below. 
Similar to LWC, expressions for $g_{es}(r)$ will be borrowed
from classical liquid-state theory; however, the particular liquid-state
theories used will differ. Those will be discussed in the next section.

Using the variational form for the free energy, and Eq.(\ref{eq:deltamu}), an
expression for $\Delta\mu$ can be obtained. It is given in LWC.\cite{laria91}
Other useful properties of the solvated electron are its average kinetic
energy, $\langle ke\rangle$, potential energy, $\langle pe\rangle$,
and polymer diameter ${\cal R}(\beta\hbar/2)$.
Expressions for all of these quantities are given in LWC,\cite{laria91} with
$\langle ke \rangle$ also in Malescio and Parrinello.\cite{malescio87}
Another useful property is the electron polymer radius of gyration,
the square of which is
\begin{eqnarray}
R_g^2\equiv && {1\over\beta\hbar}\int_0^{\beta\hbar}d\tau\
\langle ({\bf r}(\tau) - {\bf r}_{cm})^2\rangle \nonumber \\
= && 6\sum_{n=1}^\infty {1\over m\beta\Omega_n^2 + \gamma_n},
\label{eq:Rg}
\end{eqnarray}
where the last expression was derived using the reference action $S_0+S_{ref}$
above. Here, ${\bf r}_{cm}$ is the position of the polymer center of mass.
Predictions for all these properties will be given below. 

\subsection{\label{subsec:rorpa_n_drl_theory}Review of RO-RPA and DRL liquid-state theories}
In this section, two theories that give predictions for the radial distribution function in molecular liquids are reviewed briefly. These will be used as
self-consistent inputs to the hydrated electron theory through Eq.(\ref{eq:RISM}).

\subsubsection{\label{subsubsec:rorpa}RO-RPA}
In the RPA, the intermolecular correlation function $h_{es}(r)$ is represented 
 by a simple expression in terms of the intermolecular potentials and
pair intramolecular correlation functions. For the electron-solvent system it 
has the form:
\begin{equation}
\rho h_{es}(r) = -\sum_{s'}\omega_e*\beta u_{es'}*\chi_{s's}(r).
\label{eq:rpa}
\end{equation}
As can be seen, this equation is identical to the RISM equation,
Eq.(\ref{eq:RISM}), with the
substitution $-\beta u_{es}(r)\rightarrow c_{es}(r)$, the RISM equation
itself reducing
to the Ornstein-Zernike (OZ) equation\cite{hansen86} for a liquid of atoms.

As discussed in Sec. \ref{subsec:rism_polaron_theory} above, the RPA is not expected to work
well for systems with strong interactions such as that experienced by the
hydrated electron. One sign of this breakdown is that for strong repulsive
interactions, the radial distribution function will be negative at
short distances, even though it strictly is a positive quantity.

In OZ theory, for atoms interacting with potentials
that are hard-core for distances $r<\sigma$, and mildly attractive for
$r>\sigma$, one solution to the problem is to employ the mean spherical
approximation (MSA) closure.\cite{hansen86} This closure for a
single component liquid consists of demanding
that $g(r)$ be exactly zero for $r<\sigma$, and $c(r) = -\beta u(r)$ for
$r>\sigma$, where $u(r)$ is the attractive potential. In this manner, one
solves for $g(r)$ for $r>\sigma$ and $c(r)$ for $r<\sigma$.
An analogous method, the 
Optimized-RPA (ORPA), can be used to optimize the RPA free energy.\cite{andersen71}
The MSA closure has found much use in RISM and polymer RISM theory,\cite{schweizer97} and has been shown
to be diagrammatically proper for the polymer version\cite{melen97} of the
theory of Chandler, Silbey and Ladanyi.\cite{chandler82} 

The limitation to this approach though is that if particles interact
with Coulomb forces, the long-ranged potential may itself produce strong
repulsion. In this case, the OZ-MSA or ORPA method will still yield a $g(r)$
that is negative for distances outside the hard-core range $\sigma$.

A remedy is to notice that a strongly repulsive Coulomb
 interaction, as far as the radial distribution function is concerned,
produces essentially the same effect as a hard-core
potential, namely making $g(r)$ very close to zero at small (or not so small
for polyelectrolytes) $r$.\cite{donley04a} Given
that, replace the potential $u(r)$ by an optimized one, 
${\tilde u}(r)$, which has an effective hard-core range
of $\sigma^{eff}$. The range is chosen to have the smallest
value such that $g(r)$ is positive for all $r>\sigma^{eff}$,
subject to the constraint that
$\sigma^{eff} \geq \sigma$. As for the MSA, inside the effective hard-core,
${\tilde u}(r)$ is determined by enforcing the constraint $g(r)=0$, and
outside, ${\tilde u}(r)$ equals $u(r)$.
For the electron-solvent case, each potential $u_{es}(r)$ in
Eq. (\ref{eq:rpa}) is
replaced by an optimized one ${\tilde u}_{es}(r)$ with their own effective
hard-core diameters $\sigma_{es}^{eff}$.

This RO-RPA theory has been shown to give predictions for the static 
structure factor and osmotic pressure of strongly charged polyelectrolyte
solutions that are in quantitative agreement with simulation and
experiment.\cite{donley06}

\subsubsection{\label{subsubsec:drl_theory}DRL}
Another successful approach toward understanding the structure of molecular liquids is to notice that, as an electron in solvent can be represented as a
single electron in an effective field, so can the pair correlations between
two molecules be represented as a configurational average of these two molecules
in an effective field.\footnote{In classical statistical mechanics any $N$-site
correlation function can be represented {\it exactly} as the configurational
average of $M\leq N$ molecules in an effective field.}
Following this idea leads
one to the two-chain equation for the radial distribution function. The equation was first suggested by Chandler and co-workers,\cite{chandler86,laria91}
and later derived by Donley, Curro and McCoy\cite{donley94} using CMS density
functional theory.

For electron-solvent correlations, the two-chain equation is
\begin{eqnarray}
g_{es}(r) = {1\over \beta\hbar}\int_0^{\beta\hbar}d\tau  
\Bigl\langle && exp\bigl[-V_{eff}([\bfr(\tau^\prime)],\{\bfR_{1}\})\bigr]
\times \nonumber \\
&& V\delta(\bfr - \bfr(\tau) + \bfR_{1s})\Bigr\rangle,
\label{eq:two_chain}
\end{eqnarray}
where the brackets denote a configurational average of the electron polymer and
solvent molecule 1, $\{\bfR_1\}$ denotes the set of coordinates of that solvent molecule, and
the effective electron-solvent interaction potential is
\begin{equation}
V_{eff}([\bfr(\tau^\prime)],\{\bfR_{1}\}) = {1\over \beta\hbar}\int_0^{\beta\hbar}d\tau^\prime\ \sum_{s'}v_{es'}^{eff}(\bfr(\tau^\prime) -\bfR_{1s'}).
\label{eq:Veff}
\end{equation}
Note the similarity
of Eq.(\ref{eq:two_chain}) with Eq.(\ref{eq:gofr_def}).

The effective pair electron-solvent interaction potential in Eq.(\ref{eq:Veff}) is
\begin{equation}
v_{es}^{eff}(r) = \beta u_{es}(r) + v_{es}(r),
\label{eq:v_eff}
\end{equation}
where the electron-solvent medium-induced potential has an HNC form:\footnote{The HNC form of the medium-induced potential was derived. Other forms of the potential as ansatzes have been and can be explored, the only restriction being that they be pairwise decomposable.}
\begin{equation}
v_{es}(r) = -\sum_{tt'}c_{et}*\chi_{tt'}*c_{t's}(r),
\label{eq:v_mip_ew}
\end{equation}
with $c_{es}(r)$ and $\chi_{ss'}(r)$ being defined by Eqs.(\ref{eq:RISM}) and (\ref{eq:chi}),
respectively. The solvent direct correlation function $c_{ss'}(r)$ is defined
by a RISM equation similar to Eq.(\ref{eq:RISM}).

For the solvated electron, LWC approximated Eq.(\ref{eq:two_chain}) as
\begin{equation}
g_{es}(r) \approx exp\bigl[-\sum_{s'}\omega_e*v_{es'}^{eff}*\omega_{s's}(r)\bigr],
\label{eq:lwc}
\end{equation}
where the solvent intramolecular correlation function
\begin{equation}
\omega_{ss'}(r) = \bigl\langle\delta(\bfr-\bfR_{1s}+\bfR_{1s'})\bigr\rangle,
\label{eq:omega}
\end{equation}
with the brackets denoting an ensemble average over the solvent.

While giving predictions for the hydrated electron size and chemical potential
that were in reasonable agreement with simulation, LWC thought that the weakest
element of their
theory was Eq.(\ref{eq:lwc}). One reason is that molecular averaging
over the effective pair potential, Eq.(\ref{eq:v_eff}), cuts off the water
Coulomb potential at the size of the molecule, water being charge neutral.
In that manner angular correlations between the electron and the individual
water atoms beyond the molecule size are lost, reducing the polarization 
effects of the water.\cite{laria91}

A further issue brought to light in the use of Eq.(\ref{eq:lwc}) for
polyelectrolytes is that it exaggerates the repulsion between like charged
molecules due to the neglect of site level correlations.\cite{donley05}
We have investigated this effect for the hydrated electron by
solving Eq.(\ref{eq:two_chain}) by simulation using the ``cloud'' model.
In this model, the electron polymer is replaced by a single site, but to retain
some effects of the smearing of the electron charge, the electron-water
potential is replaced by $\omega_e*u_{es}(r)$. It was found that
the predictions for
$g_{es}(r)$ were very close to those of Eq.(\ref{eq:lwc}) indicating
that electron site-level correlations are important even for a compact electron
polymer.

As such, it seems reasonable to include these site level correlations in
some way. An approximation to the two-chain equation by Donley, Rajasekaran
and Liu (DRL) does this.\cite{donley98,donley02,donley04b} For the electron-solvent
system it is as follows.

Define a function $x_{es}(r,\lambda)$ for which the variable $\lambda$ is
a measure of the strength of the effective interaction, Eq.(\ref{eq:v_eff}).
The function has $\lambda$ endpoints that are powers of the radial distribution
function: $x_{es}(r,1)=g_{es}(r)^{\eta}$ and $x_{es}(r,0)=g_{es}^{ref}(r)^\eta$,
where $g_{es}^{ref}(r)$ is for a reference system, which could be a hard-core
one. The DRL equation then is:
\begin{equation}
x_{es}(r,\lambda) = x_{es}(r,0) - \eta\sum_{s'}\omega_e*\Gamma_{es'}(\lambda)*\omega_{s's}(r),
\label{eq:drl}
\end{equation}
where
\begin{equation}
\Gamma_{es}(r,\lambda) = \delta v_{es}^{eff}(r)\int_0^\lambda d\lambda^\prime\ x_{es}(r,\lambda^\prime),
\label{eq:charging}
\end{equation}
and
\begin{equation}
\delta v_{es}^{eff}(r) = v_{es}^{eff}(r) - v_{es}^{eff\_ref}(r).
\label{eq:deltav}
\end{equation}
The charging integral in $\Gamma_{es}(r,\lambda)$ is performed with
the effective interaction $\delta v_{es}^{eff}(r)$ held constant.
The term $v_{es}^{eff\_ref}(r)$ is the value of the effective potential for
the reference system. The exponent $\eta = 1/2$.

A further approximation is to assume that $x_{es}(r,\lambda)$
in $\Gamma_{es}(r,\lambda)$ varies with $\lambda$ as a simple power:\cite{donley02}
\begin{equation}
x_{es}(r,\lambda)\approx x_{es}(r,0)^{1-\lambda}x_{es}(r,1)^{\lambda}.
\label{eq:x_lambda}
\end{equation}
This form allows the charging integral to be computed analytically.
The accuracy of this approximation was determined by also solving the theory
by computing the charging integral $\Gamma_{es}(r,\lambda)$ directly.
This was done by discretizing $x_{es}(r,\lambda)$ in $\lambda$ and solving the 
resultant difference equations derived from Eqs.(\ref{eq:drl}) and (\ref{eq:charging}) above.
It was found that the results for $\Delta\mu$ and $R_g$ differed by at most
a few percent in the cases examined. As such, only results using 
Eq.(\ref{eq:x_lambda}) will be shown here.

Given the potentials $u_{es}(r)$, intramolecular structure functions
$\omega_{ss'}(r)$ and $\omega_e(r)$, and solvent correlations embodied in
$\chi_{ss'}(r)$ (and thus solvent direct correlations $c_{ss'}(r)$),
Eqs. (\ref{eq:RISM}), (\ref{eq:v_eff}), (\ref{eq:v_mip_ew}), and (\ref{eq:drl})-(\ref{eq:x_lambda}) form a closed set that can be solved numerically. The procedure
to do so will be described in Sec. \ref{sec:numerics} below.

\section{\label{sec:other_ingredients}Other ingredients}

\subsection{\label{subsec:water_sim}Water structure functions}
As mentioned in Sec. \ref{sec:theory}, the water density-density correlation
function, $\chi_{ss'}(r)$, is a necessary input to the theory.
In LWC, an MD simulation of the single point charge (SPC) water model
was used to determine the local structure of the water. Then the long
wavelength behavior of $\chi_{ss'}(r)$ was corrected.\cite{laria91}

The method here was very similar, but instead of the SPC model, the
extended SPC (SPC/E) was used. The SPC/E model has been shown to give more
realistic local correlations, pressure and dielectric constant than the
SPC.\cite{vanderspoel98,wu06}

The MD simulation was conducted using the LAMMPS package for classical 
systems.\cite{lammps} Periodic boundary conditions, a particle-particle 
particle-mesh solver and a Nose-Hoover (constant NVT) thermostat were used.
The water molecule bond lengths were held fixed using the SHAKE algorithm.
Details of the SPC/E potentials are given elsewhere.\cite{vanderspoel98}
The simulation consisted of $2\times 10^4$ water molecules at a
temperature of 298 K and density of $0.997\ {\rm g/cm^3}$. This gave a box
length $L\simeq 84.347$\AA. Integration of the
equations of motion was done with a timestep of 2 fs and the simulation was run
to 20 ns. The initial state was random and equilibrium was reached
at most by 4 ns. The equilibrium correlations functions $g_{ss'}(r)$, $s=O$ and
$H$, were then computed using data for times greater than 4 ns.

The partial water structure functions $\chi_{ss'}(r)$, Eq.(\ref{eq:chi}), were
computed using the standard relation:\cite{hansen86}
\begin{equation}
\chi_{ss'}(r) = \rho \omega_{ss'}(r) + \rho^2 h_{ss'}(r),
\label{eq:chi_hat}
\end{equation}
where $h_{ss'}(r)=g_{ss'}(r)-1$,
and the intramolecular structure function, Eq.(\ref{eq:omega}), was computed
analytically in the SPC/E model.

It is important to the theory to model accurately the correlation functions
at small wavevector.\cite{laria91}
To increase this accuracy then, the Fourier transforms of the $\chi_{ss'}(r)$,
${\hat\chi}_{ss'}(k)$, were also computed directly for $k<1\ {\rm \AA}^{-1}$
 using a set of
wavevectors appropriate to the cubic simulation box. The Fourier
transforms of the $h_{ss'}(r)$ extracted from these
${\hat\chi}_{ss'}(k)$ were then fit to a power series in $k^2$ up to
$k^6$. The $k=0$ coefficient for all factors was set to the average value of
the initial independent fits. This average gave a compressibility of
$4.55 \times 10^{-5}$ atm$^{-1}$, close to the experimental value of $4.58\times 10^{-5}$. The $k^2$ coefficients were then adjusted slightly to give a dielectric constant of
$77$.\cite{hoye76,chandler77} These series were then joined
with the data for $k\geq 1\ {\rm \AA}^{-1}$ obtained via $g_{ss'}(r)$ to give values for
${\hat\chi}_{ss'}(k)$ for all $k$. 

As discussed in the next section, one pseudopotential used, that of 
Larsen, Glover and Schwartz, has a third type of site, call it X, on the water
molecule.\cite{larsen10}
Bulk water correlations were thus also needed for it, and were obtained as
follows. In the MD simulation done here, all averages were computed afterward
using snapshots (taken at equally spaced times) of the simulation configuration.
Since the relative position of site X was known it was
then straightforward to add its coordinates for every molecule to the stored
simulation configurations. The correlation functions $g_{ss'}(r)$ and
${\hat\chi}_{ss'}(k)$ with $s,s'$ now including site X, were then computed in
the same manner as above. 

\subsection{\label{subsec:pseudopotentials}Pseudopotentials}
The last ingredient of the theory is the choice of electron-solvent potential.
Since the water is being treated classically, this potential must incorporate
quantum effects, such as the orthogonality of the wavefunction of the solvated
electron to those of the bound water electrons, in some manner.

There are currently a number of theories for the electron-water
pseudopotential.\cite{sprik86,schnitker87a,turi02,jacobson10,larsen10} Five of
these will be implemented here, all having been used or optimized within
the SPC model of water.

The first two models, denoted in LWC as model I and II, were used by
 Sprik, Impey and Klein in their
path integral Monte Carlo study of the hydrated electron.\cite{sprik86}
In terms of the Bjerrum length $\lambda_B$, the model potentials have the 
form (the definition differing slightly from that in LWC):
\begin{equation}
\beta u_{es}(r)=\begin{cases} z_ez_s\lambda_B/R_s,& r < R_s, \\ z_ez_s\lambda_B/r,& r > R_s.\end{cases}
\label{eq:model1n2}
\end{equation}
Here, $R_O=0$ for both models, and $R_H=0$ and 1 \AA\ for models I and II,
respectively. 
Also, $z_e$ and $z_s$ are the reduced charges of the electron and water site
$s$, respectively.
In the work here, the water site charges were set to be
the same as for the SPC/E model, so $z_e=-1$, $z_O=-0.8476$ and $z_H=0.4238$.
 These values
differ slightly from those of the SPC model, so it might be thought that the latter
ones should be used. However, it was found to be important that the strength
of the electron-water interactions be consistent with the strength of
the water-water interactions.

The third model, denoted here as model SR, is the same as described by Schnitker
and Rossky,\cite{schnitker87a,schnitker87b}
except that the oxygen and hydrogen charges were changed to those of the SPC/E
model. Recently, the derivation of this model was shown by Larsen, Glover and
and Schwartz to contain an error.\cite{larsen09} While the corrected model, call it model SR-C, gives predictions that are qualitatively similar to the
original,\cite{casey13} there are noticeable quantitative differences. Thus,
results of model SR-C will also be shown.

The fourth model, denoted here as model TB, is the same as described by Turi and Borgis,\cite{turi02} except that
the oxygen and hydrogen charges were changed to those of the SPC/E model.
Model TB was optimized for SPC water. It will be shown in 
Sec. \ref{sec:results} that using SPC/E values does change the predictions, 
but perhaps only slightly.

The fifth model, denoted here as model LGS, is the same as that described by
Larsen, Glover and Schwartz,\cite{larsen10} except that the oxygen and hydrogen
charges were changed to those of the SPC/E model. 
Like model TB, model LGS was optimized for SPC water.
Model LGS includes an additional site, call it X, halfway between the two
hydrogen atoms.

\subsection{\label{subsec:anissue}An issue}
An analog of this liquid-state polaron theory for hard-core polymer melts
is self-consistent polymer RISM theory.\cite{melen93,schweizer97,curro99,heine05} For that theory,
it is found that if the chain structure is determined using an HNC
form for the medium-induced potential, similar to Eq.(\ref{eq:mip2}), then at
high densities the polymer chain may collapse upon itself.\cite{melen93} 
Since a hard-core polymer in a melt should be at least as large as an ideal
chain, this behavior is unphysical. The cause is thought to be 
that this HNC
medium-induced potential over-estimates the interaction of the chain with the
surrounding medium at these high densities.\cite{melen93}
This same behavior occurs also for the electron-water system using the RO-RPA
and DRL theories for model I, and the DRL theory for model LGS, with no
numerical solution found.

A remedy for polymer melts is to weaken the
strength of the medium-induced potential.\cite{grayce94,donley94,heine05} 
One way is to develop molecular analogs of the Percus-Yevick\cite{hansen86} and
Martynov-Sarkisov\cite{martynov83} closures of atomic liquid-state theory, which have been
found to give weaker medium-induced interactions than the HNC at high 
density.\cite{grayce94,donley94} A variation of this method will be 
implemented here. Define a new medium-induced potential,
\begin{equation}
v^*(r) = -sgn(w)\vert w\vert^{1/\zeta} + 1,
\label{eq:v_mip_star}
\end{equation}
where
\begin{equation}
w = 1-\zeta v(r),
\label{eq:w}
\end{equation}
with $v(r)$ given by Eq.(\ref{eq:mip2}) and the exponent $\zeta \geq 1$.
By changing the value of $\zeta$, $v^*(r)$ interpolates between 
an HNC form ($\zeta=1$) and a Martynov-Sarkisov one ($\zeta=2$).\footnote{It
was not found necessary to weaken the potential even further to attain
an effective Percus-Yevick form.}
Also, as $\rho\rightarrow 0$, $v^*(r)$
reduces to an HNC form as required. For the hydrated electron, $\zeta$
will be chosen as the smallest value ($\geq 1$) needed to
prevent the chain from collapsing.
It is found for model I that $\zeta$ = 1.0173 and 1.0176 for the RO-RPA
and DRL-based theories, respectively, and $\zeta=1.0223$ for DRL for model LGS.

Now, the size of the electron ring is determined by a balance between the
pressure of the surrounding medium acting to collapse the ring, and the ring
kinetic energy acting to expand it. The scheme above though assumes it is the pressure
of the surrounding medium that is being overestimated and corrects for that.
However, Sumi and Sekino\cite{sumi04} and others\cite{wuThesis} have
offered evidence that the effective action, Eq.(\ref{eq:Sref}), tends to 
underestimate the kinetic energy of the electron. Further evidence for this
underestimation in given in Sec. \ref{sec:results} below. It is possible then,
especially as $\zeta$ is always very close to one, that this collapse
of the electron is due to that rather than from the medium-induced
potential having an HNC form. So, while the above corrective scheme will
be used in this work, no specific cause should be inferred from it.

For the LWC expression for $g_{es}(r)$, Eq.(\ref{eq:lwc}), LWC found that 
for values of $R_s$ at and near that used in model II, the polaron theory
predicted a
``super-trapped'' electron polymer of size $\simeq 1.3$ \AA.\cite{laria91}
This structure would appear to arise from the limitations of
Eq.(\ref{eq:lwc}) that favor a second, normally higher energy, solution of
the theory. This solution can be removed by the same method as above though,
and will be done so here. For this case, $\zeta = 1.0173$. With the SPC/E
water structure factors used here, it was found that model I also
gave a super-trapped polymer. This solution was removed by setting
$\zeta = 1.002$.

\section{\label{sec:numerics}Numerical solution}
All functions were solved on a grid of $N_r = 2^{11}=2048$ points. The maximum
value of $r$, $r_{max}$ was set to $\simeq 42.2$ \AA, which was half the water
simulation box length $L$, described in Sec. \ref{subsec:water_sim} above.
Since the water correlation functions were fit at low $k$, $r_{max}$ could
have been set to be much larger; however, it was found that the results did
not change for larger values.  The real space grid spacing
$\Delta r=r_{max}/N_r\simeq 0.0206$ \AA, with a reciprocal grid spacing $\Delta k = \pi/r_{max}$.
The water density and temperature were the same as for
the water simulation, so the Bjerrum length $\lambda_B = 560.8$ \AA.

The number $N_n$ of variational mode amplitudes $\gamma_n$ was set to 500, and
the imaginary time integrals were computed on a grid of $N_\tau=200$ points.
However, tricks were done to take implicitly any sum over $n$ to infinity
and integrate the time integral near $\tau=0$ and $\beta\hbar$ very accurately.
Details are in Nichols et al.\cite{nichols84}

The theory was solved in a manner similar to that described in
Nichols et al. and LWC.
First, the initial value for ${\hat\omega}_e(k)$ was taken to be its free
electron form,\cite{laria91} so the mode amplitudes $\gamma_n$ were initially
set to zero. Initial values for
$-\beta{\tilde u}_{es}(k)$ (RO-RPA) or ${\hat c}_{es}(k)$ (DRL) were set to be
$-\beta {\hat u}_{es}(k)$, the Fourier transform of the electron-water
pseudopotentials. Given these initial values and the water-water
structure factors, ${\hat\chi}_{ss'}(k)$, the RO-RPA or DRL theory was solved
for $g_{es}(r)$, and $\beta{\tilde u}_{es}(k)$ or ${\hat c}_{es}(k)$. The 
procedure to solve the RO-RPA theory is given elsewhere.\cite{donley05} The 
procedure to solve the DRL theory differed slightly from that described in
previous work\cite{donley02} and is given in the Appendix. 

With $\beta{\tilde u}_{es}(k)$ or ${\hat c}_{es}(k)$, and ${\hat\chi}_{ss'}(k)$,
the medium-induced potential, ${\hat v}^*(k)$,
was obtained using Eqs.(\ref{eq:mip2}), (\ref{eq:v_mip_star}) and (\ref{eq:w}).
Except for the cases specified in Sec.\ref{subsec:anissue} above, $\zeta$ was
set to 1. With this medium-induced
potential and the guess for ${\hat\omega}_e(k)$, new values for
the $\gamma_n$ were determined using Eq.(\ref{eq:gamman}). With these
$\gamma_n$, a new
value for ${\hat\omega}_e(k)$ was determined using the action
$S_0+S_{ref}$.

These equations were iterated until convergence was obtained. The allowed
error tolerance was $10^{-4}$ for ${\hat c}_{es}(k)$ for all $k$ and $\gamma_n$
for all $n$.
For ${\hat\omega}_e(k)$, its old and updated solutions were mixed in a ratio of
1:1 to produce a new guess. The mixing values for the RO-RPA and DRL theories
are given in the Appendix.

All grid point or sum numbers, $N_r$, $N_n$ and $N_\tau$ were increased to 
determine if the results varied, but they did not. 

\section{\label{sec:results}Results}
First, as mentioned above, the radial distribution function $g_{es}(r)$ computed
by the theory represents correlations between a site on the electron
ring and a water site. In contrast, most published work from mixed
quantum/classical simulations of the hydrated electron show the electron-water
correlations in terms of the electron center-of-mass (eCOM) and a water site.
Because the eCOM is shielded by its ring sites, eCOM-water correlations
tend to have a local structure typical of molecular liquids interacting
with harsh short range repulsive forces. The site-site
$g_{es}(r)$ usually reflects a softer interaction, especially as the single
electron charge is distributed evenly along the ring.\cite{sprik86,schnitker87b}

\begin{figure}
\includegraphics[scale=0.55]{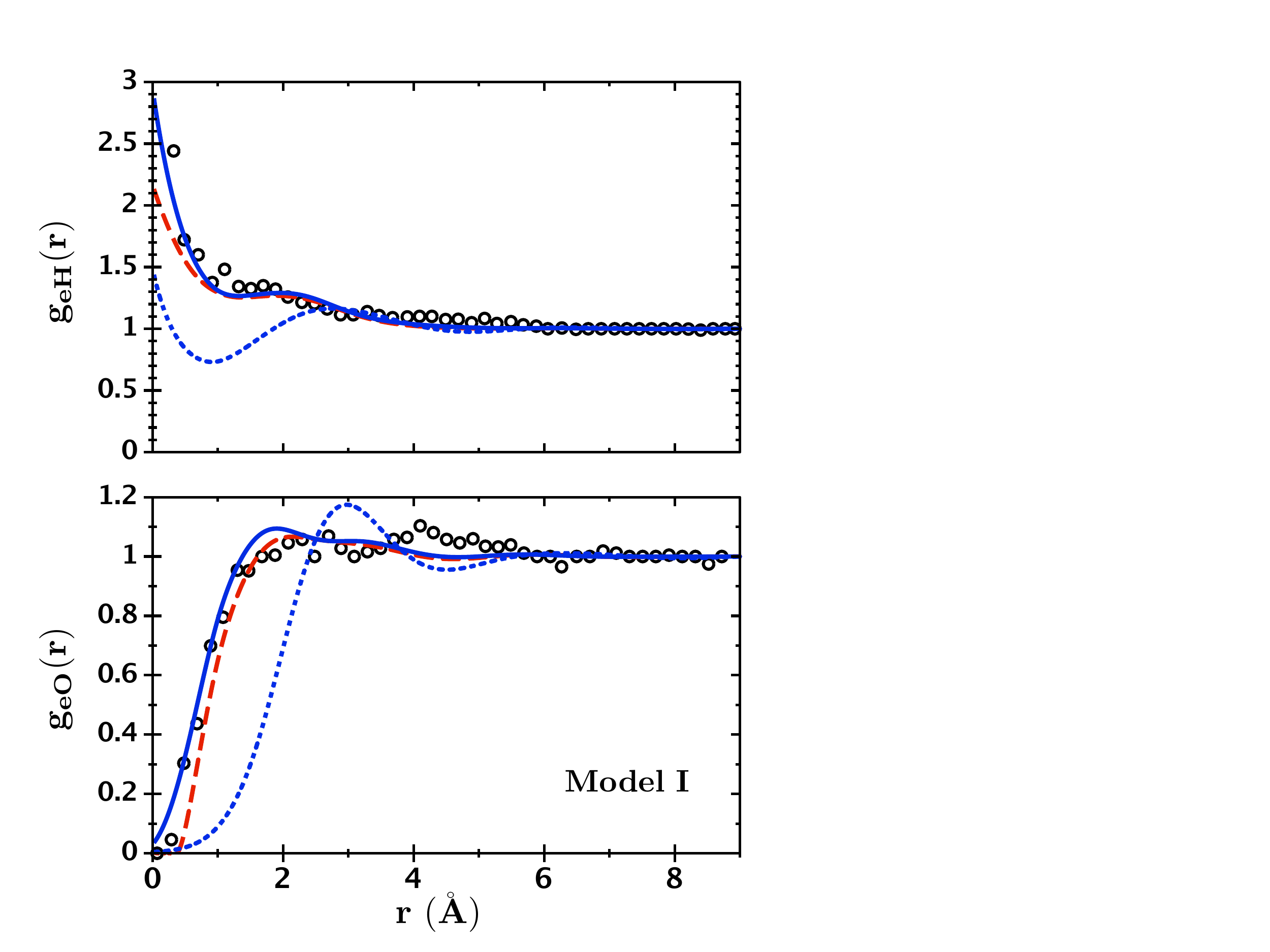}
\caption{\label{fig:grModel1}
Electron-water site-site radial distribution functions for pseudopotential
model I. The blue solid, red dashed and blue dotted lines correspond to
theoretical results using the liquid-state equations of DRL, RO-RPA and LWC,
respectively.
The black circles are simulation data of Sprik et al.\cite{sprik86}
}\end{figure}

\begin{figure}
\includegraphics[scale=0.55]{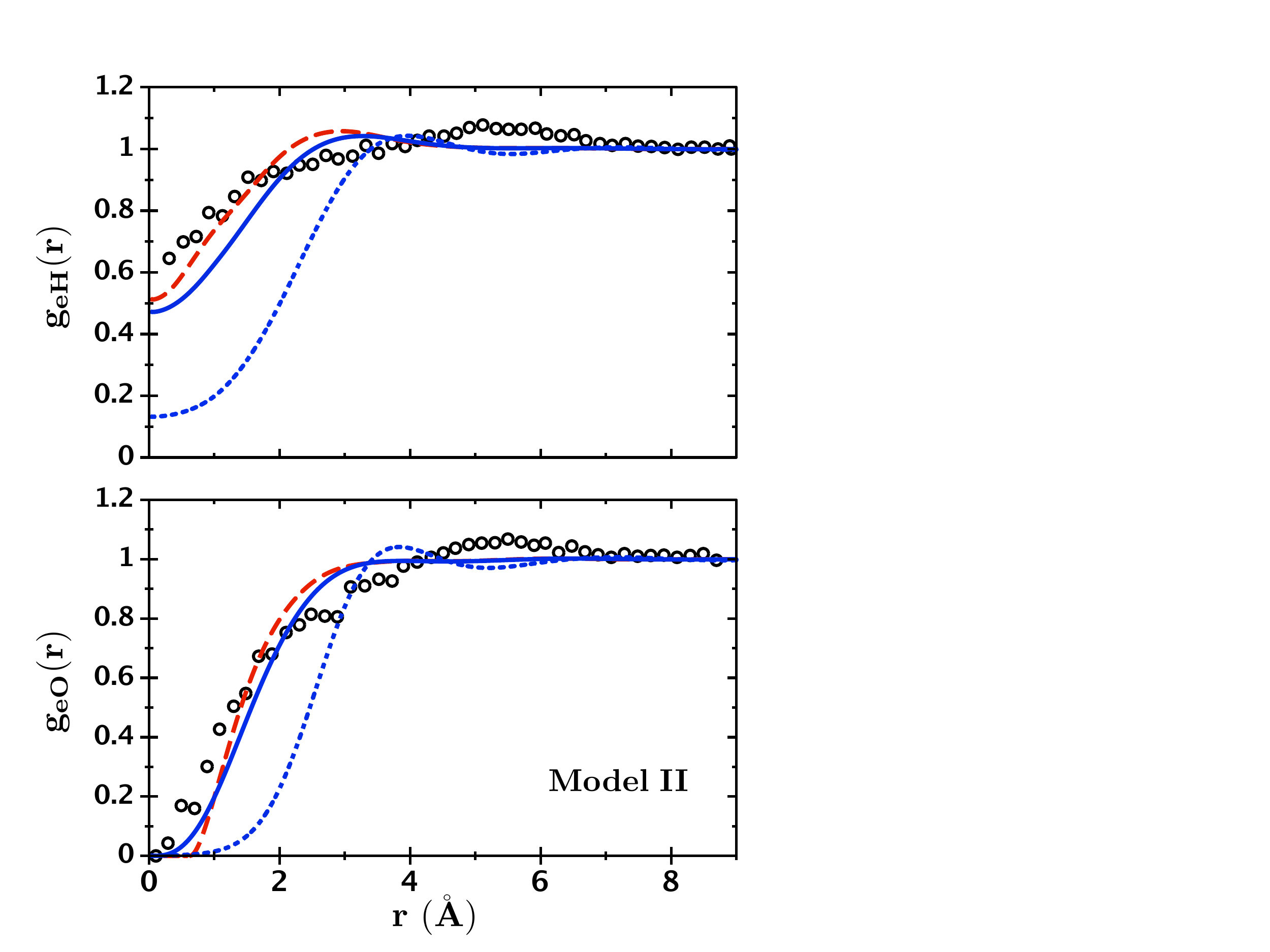}
\caption{\label{fig:grModel2}
Electron-water site-site radial distribution functions for pseudopotential
model II. The blue solid, red dashed and blue dotted lines correspond to
theoretical results using the liquid-state equations of DRL, RO-RPA and LWC,
respectively.
The black circles are simulation data of Sprik et al.\cite{sprik86}
}\end{figure}

Figures \ref{fig:grModel1} and \ref{fig:grModel2} show results for $g_{es}(r)$
using the pseudopotentials of models I and II, respectively. Polaron 
theory predictions using the LWC, RO-RPA and DRL equations for $g_{es}(r)$
are shown, along with path integral Monte Carlo simulation data of Sprik, Impey
and Klein.\cite{sprik86} As can be seen, the agreement between theory and
simulation is much improved using the RO-RPA and DRL equations for $g_{es}(r)$
in comparison to that of LWC, especially at small $r$. Differences remain
though. The RO-RPA and DRL-based theories overestimate $g_{es}(r)$ at
$r\sim 2.5$ \AA\ for model II.
This disagreement was investigated and found to be due to
using the SPC/E model to compute ${\hat\chi}_{ss'}(k)$, this model producing
stronger local correlations than the SPC.
This effect can be seen most easily by comparing the LWC-based results for
$g_{eO}(r)$ for model I and II in the present work with those in the original
LWC paper, which used water structure factors computed from an
MD simulation of the SPC model. The solvation ordering for $g_{eO}(r)$
is more pronounced in the present work. 

Table \ref{table:energies} shows predictions of the electron polymer
diameter ${\cal R}(\beta\hbar/2)$ and radius of gyration $R_g$. For models I and
II, these are within 8\% and 5\%, respectively, of the simulation values for all
theories. There are no simulation values for the excess chemical potential
$\Delta\mu$, but the predictions of the RO-RPA and DRL-based theories are
within 27\% of the experimental value for both models.

It can also be seen from Table \ref{table:energies} that the predictions for
model I from the RO-RPA and DRL theories for
$\langle ke\rangle$ and $\langle pe\rangle$
are 35\% and 22\% less, respectively, than the simulation values.
Also, the pure Coulomb virial relation, $\langle ke\rangle = -{1\over 2}\langle pe\rangle$, is clearly not obeyed either.
As mentioned above, Sumi and Sekino have offered evidence
that the polaron theory, using the effective interaction action $S_{ref}$,
underestimates the electron kinetic energy,\cite{sumi04} so that may be
the primary cause of the discrepancy here.

The simulation data shows slightly larger electron-water correlations than
the theories at intermediate distances, $r\sim 4$ \AA\ for model I, and 
$r\sim 5$ \AA\ for model II. This enrichment would seem to be an indication
of solvation ordering. Liquid-state theories of atoms and polymers typically
are very good at
predicting such local ordering.\cite{hansen86,schweizer97,heine05} However, the hydrated electron differs from an ion or polymer (at short
lengthscales) in that the electron shape fluctuates greatly from the slow
(on electron timescales) changes in the local polarization of its 
environment. 
As discussed in Sec. \ref{subsec:rism_polaron_theory}, reducing the electron-solvent system to a single electron in an HNC-like effective field decreases the
effect of this local polarization. Yet, given the 
agreement with the simulation for other properties, this difference does not
appear to be an important one.

The predictions of the RPA theory using model I were also examined. For
simplicity, the RPA was also used for the water structure factors,
${\hat\chi}_{ss'}(k)$. As can be seen in Table \ref{table:energies},
the RPA predictions for $\Delta\mu$ and $R_g$ agree very well
with experiment, even though, as expected, the RPA predicts that
$g_{eO}(r)$ is negative at small distances, as indicated by the large
negative value of $\langle pe\rangle$.

\begin{table}[h]
\caption{Theoretical values for various chain sizes and energies, and comparison
with those of simulation and experiment. All lengths are in angstroms (\AA) and all energies are in electron-volts ($eV$).}
\centering
\begin{tabular}{l c ccccc}
\hline\hline
Model & Source & ${\cal R}(\beta\hbar/2)$ & $R_g$ & $\Delta\mu$ & $\langle ke\rangle$ & $\langle pe \rangle$ \\ [0.5ex] 
\hline
 & Experiment &  & 2.5\cite{bartels05} & -1.71\cite{jortner66} &  & \\
\hline
I & RO-RPA & 2.83 & 1.99 & -1.91 & 2.15 & -8.37  \\
  & DRL    & 2.99 & 2.09 & -2.17 & 1.95 & -8.03  \\
  & LWC    & 2.61 & 1.83 & -1.71 & 2.51 & -8.17  \\
  & RPA    & 3.55 & 2.47 & -1.54 & 1.37 & -192.7 \\
  & Sprik et al.\cite{sprik86} & 2.84 & & & 3.3 & -6.7  \\
\hline
II & RO-RPA & 3.09 & 2.16 & -1.44 & 1.81 & -6.35  \\
   & DRL    & 3.17 & 2.21 & -1.29 & 1.71 & -6.06  \\
   & LWC    & 3.12 & 2.18 & -0.37 & 1.83 & -5.34  \\
   & Sprik et al.\cite{sprik86} & 3.25 & & & & \\
\hline
SR & RO-RPA & 2.88  & 2.02  & -2.85 & 2.05 & -8.46 \\
    & DRL   & 2.97  & 2.08  & -2.70 & 1.94 & -8.00 \\ 
    & Schnitker et al.\cite{schnitker87b} & & 2.1 & -2.2 & & \\
\hline
SR-C & RO-RPA & 3.56  & 2.48  & -0.83 & 1.39 & -4.50 \\
          & DRL    & 3.47  & 2.42  & -0.72 & 1.46 & -4.49 \\ 
    & Larsen et al.\cite{larsen09} & & 2.6 & & & \\
\hline
TB & RO-RPA & 3.06  & 2.14  & -2.11 & 1.83 & -7.08 \\
   & DRL    & 3.16  & 2.21  & -1.96 & 1.71 & -6.73 \\
   & Turi et al.\cite{turi02} & & 2.42 & & & \\
\hline
LGS & RO-RPA & 3.74  & 2.60  & -5.39 & 1.27 & -8.99 \\
    & DRL    & 2.96  & 2.07  & -7.66 & 1.99 & -11.26 \\
   & Larsen et al.\cite{larsen10} & & 2.6 & & & \\
\hline
\end{tabular}
\label{table:energies}
\end{table}

The electron-oxygen repulsion in pseudopotential models SR and SR-C
is large at short distances. Though not necessary,
it was helpful then to solve the DRL equation for this model
with respect to a hard-core reference $g_{es}^{ref}(r)$. The RISM\cite{chandler72} predictions
for $g_{ss'}^{ref}(r)$ and $g_{es}^{ref}(r)$ were used.
For model SR, the hard-core diameters were set to be $\sigma_{OO}=2.4$ \AA, $\sigma_{OH}=1.4$ \AA, $\sigma_{HH}=1.4$ \AA, $\sigma_{eO}=0.8$ \AA\ and $\sigma_{eH}=0.0$ \AA.
For model SR-C the same values were used, except $\sigma_{eO}=1.0$ \AA\ and
$\sigma_{eH}=0.5$ \AA.

\begin{figure}
\includegraphics[scale=0.55]{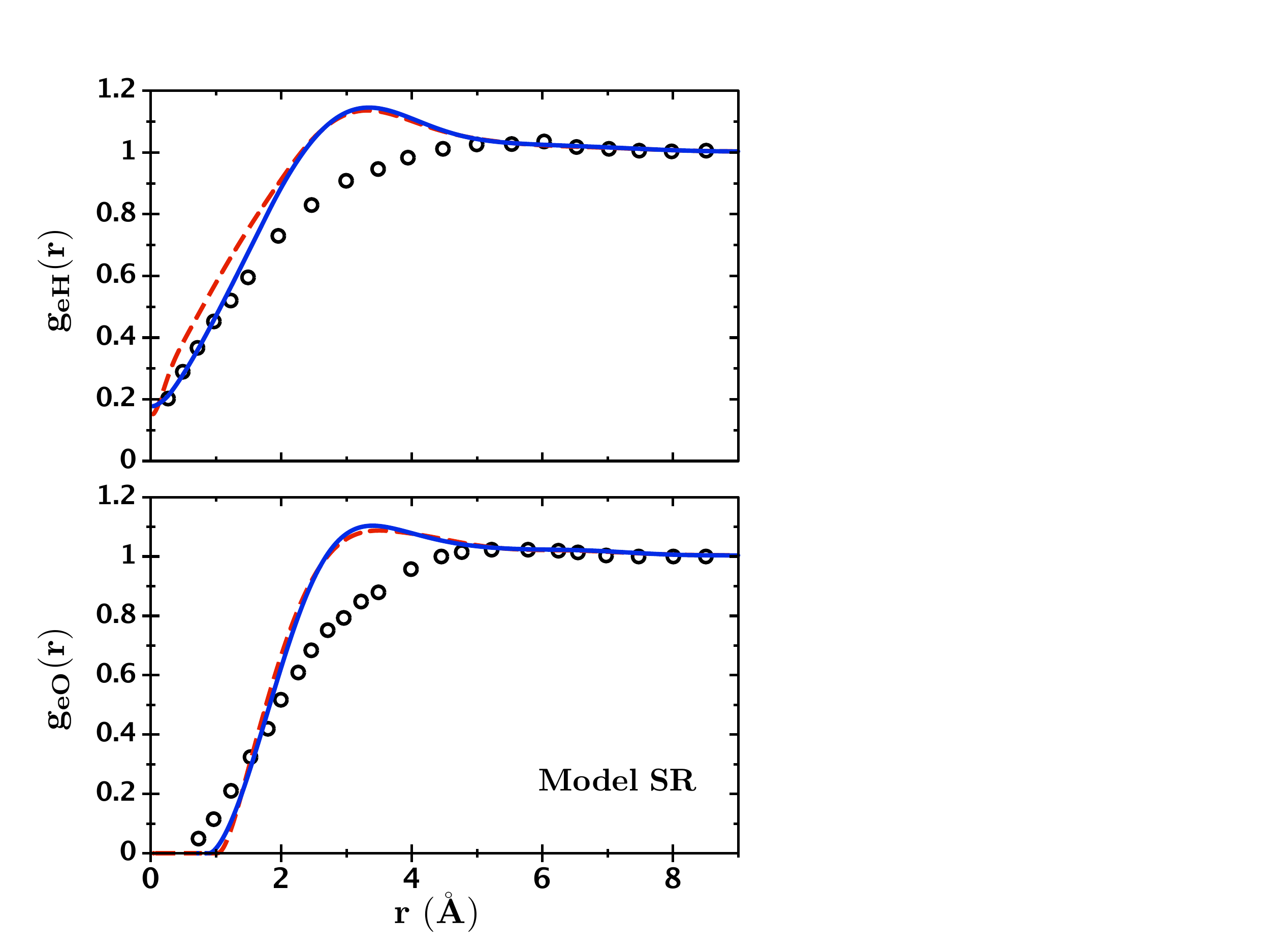}
\caption{\label{fig:grModelSR}
Electron-water site-site radial distribution functions for pseudopotential
model SR. The blue solid and red dashed lines correspond to
theoretical results using the liquid-state equations of DRL and RO-RPA,
respectively.
The black circles are simulation data of Schnitker and Rossky.\cite{schnitker87b}
}\end{figure}

Figure \ref{fig:grModelSR} shows results for $g_{es}(r)$ using model SR.
As for models I and II, the RO-RPA and DRL-based polaron theories give very
similar predictions. The theories predict modest peaks at
$r\sim 3$ \AA\ for both $g_{eO}(r)$ and $g_{eH}(r)$, which
are not present in the simulation data.\cite{schnitker87b} The cause appears to be the same as
the over-estimation of $g_{es}(r)$ at $r\sim 2.5$ \AA\ for model II, namely
the use of the SPC/E water model instead of the SPC to compute the bulk
water correlations. Differences in the treatment of the long-range interactions
in the MD simulation here and as done by Schnitker and Rossky may also
contribute.
The theories agree well with the SR data at other distances $r$ though.

As shown in Table \ref{table:energies} the theoretical values for $R_g$ for
both theories agrees very well with that from simulation, within 4\%.\cite{schnitker87b}
The predictions for $\Delta\mu$ are less accurate, being 30\% and 23\%
more negative than the SR estimate for the RO-RPA and DRL-based
theories, respectively.
Properties using the corrected Schnitker-Rossky
pseudopotential, model SR-C, are also shown in the table.

\begin{figure}
\includegraphics[scale=0.55]{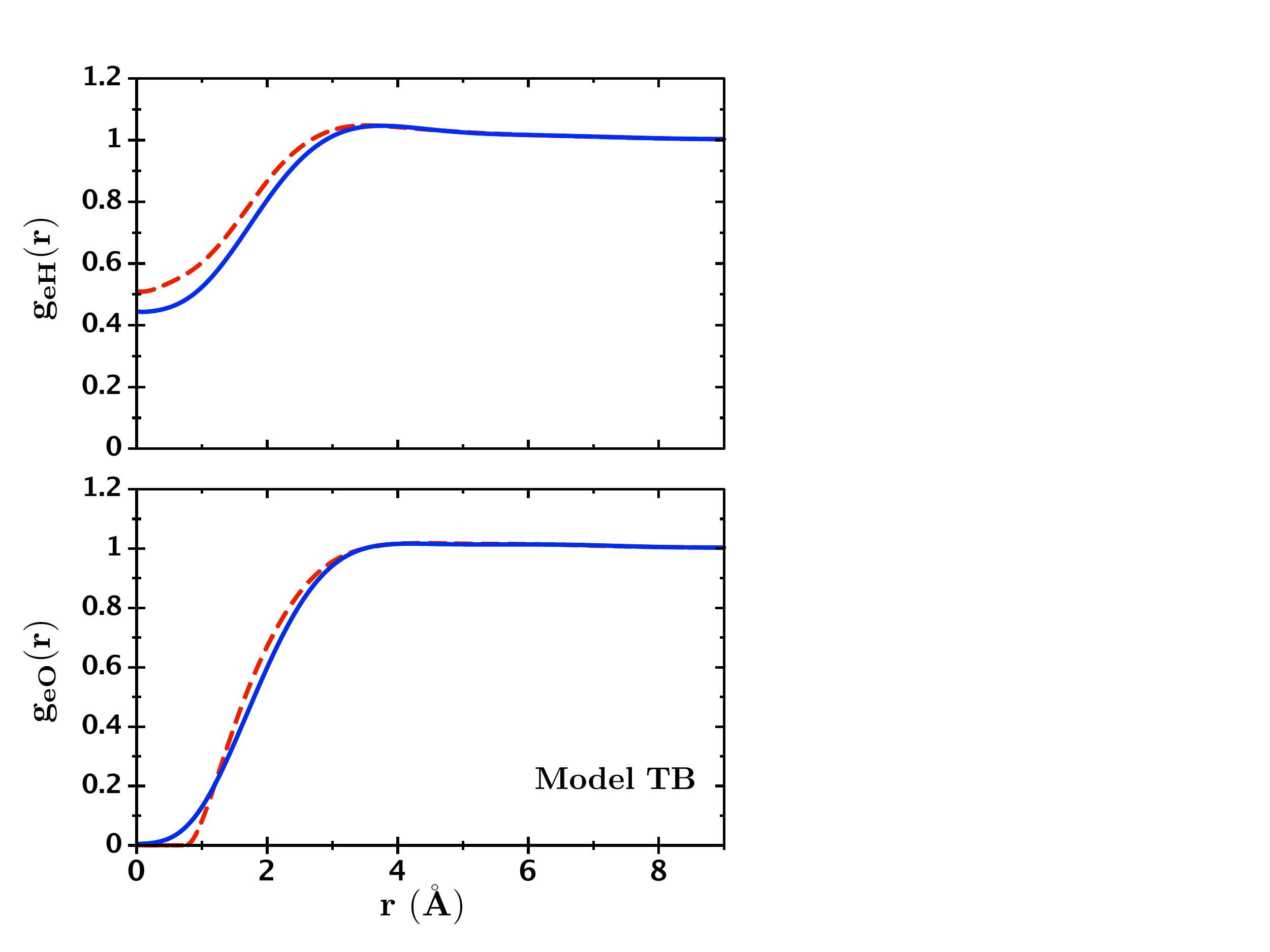}
\caption{\label{fig:grModelTB}
Electron-water site-site radial distribution functions for pseudopotential
model TB. The blue solid and red dashed lines correspond to
theoretical results using the liquid-state equations of DRL and RO-RPA,
respectively.
}\end{figure}

Figure \ref{fig:grModelTB} shows results for $g_{es}(r)$ using model TB.
As for the other pseudopotentials, the
RO-RPA and DRL-based polaron theories give very similar predictions. Model TB
was designed to give greater penetration of the hydrogen cloud by the hydrated 
electron than the Schnitker-Rossky potential.\cite{turi02} Comparing
$g_{eH}(r)$ in Figures
\ref{fig:grModelSR} and \ref{fig:grModelTB}, it appears that Turi and Borgis
have been successful.
There is no published simulation data for the site-site radial distribution
function for model TB.

Table \ref{table:energies} shows that the values of $R_g$ predicted by the
RO-RPA and DRL theories are less than the simulation value computed by Turi
and Borgis\cite{turi02} by 12\% and 9\%, respectively. However, as was
 mentioned above,
model TB was optimized for SPC water, not SPC/E.
RO-RPA and DRL predictions for $\Delta\mu$ with model TB are more negative than
the experimental value by 23\% and 15\%, respectively.

\begin{figure}
\includegraphics[scale=0.75]{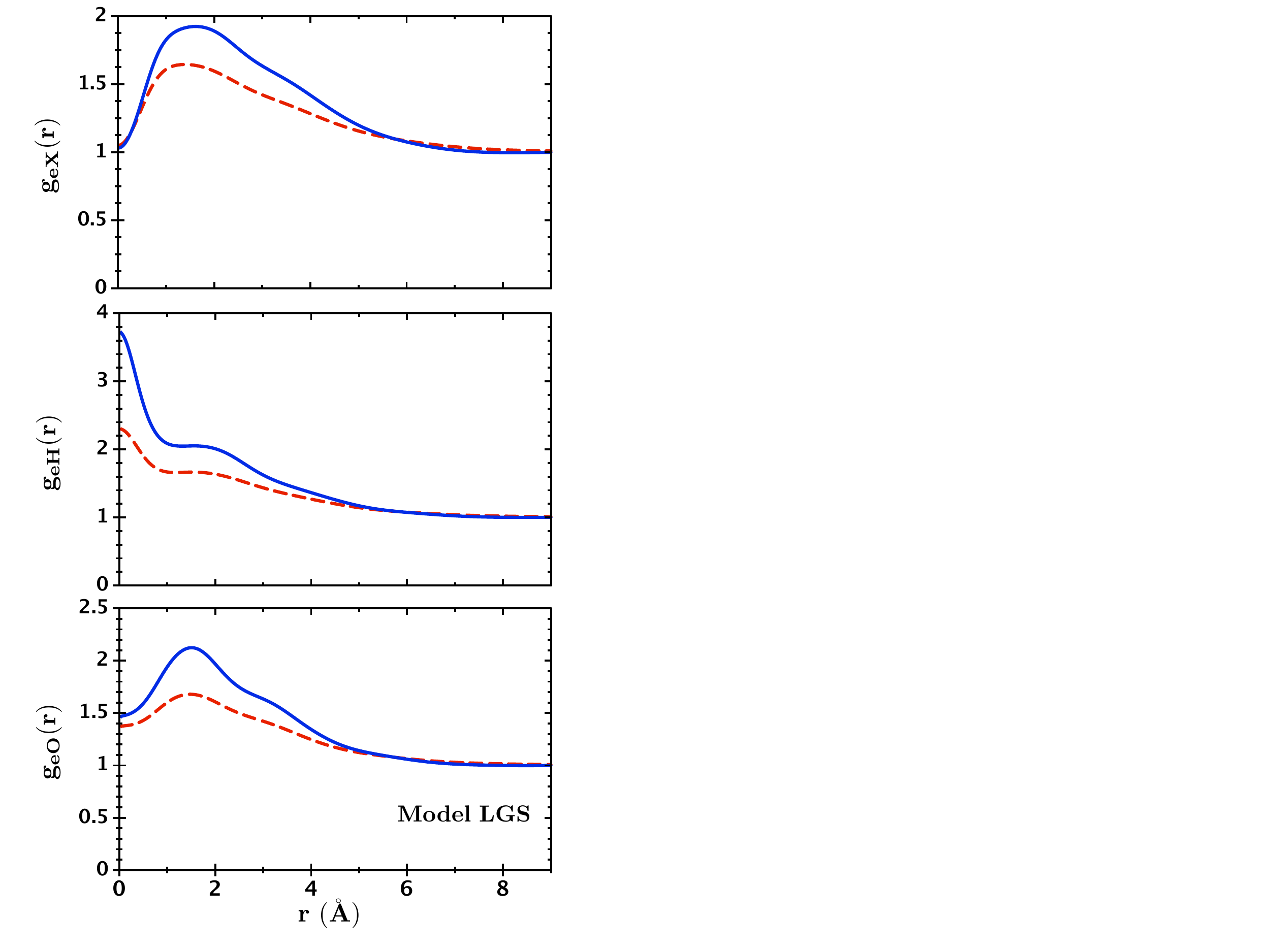}
\caption{\label{fig:grModelLGS}
Electron-water site-site radial distribution functions for pseudopotential
model LGS. The blue solid and red dashed lines correspond to
theoretical results using the liquid-state equations of DRL and RO-RPA,
respectively. Site ``X" corresponds to the extra site on the water
molecule.
}\end{figure}

Figure \ref{fig:grModelLGS} shows results for $g_{es}(r)$ using the Larsen,
Glover and Schwartz pseudopotential, model LGS. The RO-RPA and DRL-based
theories are in qualitative agreement for all correlations. Both show that
the water density is increased near the electron,\cite{larsen10,turi12} with
even $g_{eO}(r)$ above unity for all distances $r<8$ \AA. The hydrated 
electron seems to be acting as a source of cohesion for the water in a manner
similar to valence electrons in a metal.
Contrary to results
using the other pseudopotentials though, the RO-RPA and DRL-based theories are
not in close quantitative agreement. The reason for this is that while the
RO-RPA theory greatly improves correlations between sites that are repulsive,
its predictions for correlations between sites that are attractive are only
slightly better than for the RPA.\cite{donley05} Thus, it is expected that the
DRL theory gives a more accurate estimate of these enhanced correlations in
model LGS as it does for $g_{eH}(r)$ in model I.

Table \ref{table:energies} shows that the DRL-based theory predicts a value
for $R_g$ 20\% less than simulation, while the RO-RPA prediction equals the
simulation value.\footnote{Herbert and Jacobson\cite{herbert11}, 
found that employing a more accurate scheme (Ewald sum) than done by LGS
to handle the long-range interactions gave $R_g=2.7$ \AA.} This agreement
between the RO-RPA theory and simulation would seem to be due to a
cancellation of errors: it underestimates positive density correlations, which
implies an underestimation of the electron-water forces embodied in the 
medium-induced potential, and this is presumably balanced by an
underestimation of the kinetic energy in the reference action, Eq.(\ref{eq:Sref}).   

Perhaps most interestingly, both theories predict an excess chemical
potential $\Delta\mu$ that is at least three times more negative than that
from experiment. Given the degree of agreement between theory, and simulation
and experiment for the other
pseudopotential models, this result is surprising. As mentioned above, there
are elements in the theory that probably need improvement, such as a better
estimate for the electron kinetic energy. Plus, it has been shown that the use
of the SPC/E model instead of the SPC does cause noticeable changes in the
local correlations and energetics, especially if the handling of the long-range
interactions differ. Last, the LGS potential was optimized for SPC water,
and it has been shown that the LGS predictions do possess a sensitivity
above that of the other potentials.\cite{herbert11,turi12}
So, while it seems unlikely that all these aspects
combined could cause an error in $\Delta\mu$ of a factor of three, it should
still be considered as possible. 

\section{\label{sec:summary}Summary and discussion}
In summary, the RISM-polaron theory of Chandler and co-workers was improved
for the hydrated electron by using two liquid-state theories, RO-RPA and DRL,
for the electron-water correlations.
It is found that the RO-RPA and DRL-based polaron theories give similar 
results. Further, for a given pseudopotential, these
polaron theories give quantitatively accurate predictions for most static
equilibrium properties of the hydrated electron in comparison with path
integral Monte Carlo or MD simulations.

One discrepancy between theory and simulation is for the average electron
kinetic energy, $\langle ke\rangle$.
Evidence presented by others\cite{sumi04,wuThesis} appears to
point to the cause being the form of the reference action $S_{ref}$, 
Eq.(\ref{eq:Sref}), in conjunction with its determination via free energy
minimization. Given advances in computing power, a straightforward solution is
to compute the path integral directly by Monte Carlo or other means.

It was also discovered that some differences between theory and the
quantum/classical simulations were due to using different water
models. In that way, a specific electron-water pseudopotential may have meaning
only with respect to the water model(s) within which it is optimized.
Future comparisons would benefit from implementing the same water
model as simulation.

Interestingly, it was found using the LGS pseudopotential that the theories
 predict that the excess chemical potential, i.e., solvation free energy,
is less than -5 eV, that is, at least three times more
negative than the experimental value. The LGS prediction for a related quantity, the
vertical electron binding energy, has been shown to be almost twice as large
as the experimental value.\cite{herbert11} So the result here does not
stand completely alone. Nonetheless, improvements in the theory are
needed to remove uncertainties in the predictions presented here. An estimate
of the excess chemical potential using simulation data\cite{schnitker87b}
would also be helpful.

A simpler approach, using the RPA to model both the electron-water {\it and} 
water-water correlations was also examined. It was found that this RPA theory
yielded values for the electron radius of gyration and excess chemical potential that agreed well with experiment even
though the RPA predictions for local electron-water and water-water correlations
were mediocre. A similar cancellation of effects using an RPA medium-induced
potential has been shown for an analogous system, a semi-dilute solution of
polyelectrolytes.\cite{donley97}

It is noted that the theory can be extended to examine electron excited
states by performing an analog of that done in mixed quantum/classical
simulations of the hydrated electron. A similar approach has been done for the
case in which the water was modeled as a continuum dielectric.\cite{lakhno07}

\begin{acknowledgments}
We thank David Bartels for helpful correspondence.
\end{acknowledgments}

\appendix*
\section{\label{sec:appendix}Numerical solution of the DRL theory}
The DRL theory for $g_{es}(r)$ was solved as follows.
Given the water structure factors ${\hat\chi}_{ss'}(k)$, electron intramolecular
structure factor ${\hat\omega}_e(k)$, and initial guesses for the 
electron-water direct correlation functions ${\hat c}_{es}(k)$, the RISM
equation, Eq.(\ref{eq:RISM}), was solved for ${\hat h}_{es}(k)$.
These functions were then Fourier transformed to obtain $g_{es}(r)$. With these
radial distribution functions, the charging integral, Eq.(\ref{eq:charging}),
was computed using the approximation of Eq.(\ref{eq:x_lambda}). New values
for $x_{es}(r,1)$ and thus ${\hat h}_{es}(k)$ were then obtained using 
Eq.(\ref{eq:drl}). 

Define the difference between the nonconverged values of ${\hat h}_{es}(k)$
obtained from
the two-chain (or approximation thereof) and RISM equations as
$\Delta {\hat h}_{es}(k)$. Also, define the difference
between the new and old solution for ${\hat c}_{es}(k)$ as 
$\Delta {\hat c}_{es}(k)$. It can be shown for the LWC approximation 
Eq.(\ref{eq:lwc}) to the two-chain equation Eq.(\ref{eq:two_chain}) that
the Newton-Raphson solution algorithm gives:
\begin{equation}
\Delta {\hat c}_{es}(k) \approx \sum_{s'}{\hat\omega}_e^{-1}(k)\Delta {\hat h}_{es'}(k){\hat\omega}_{s's}^{-1}(k),
\label{eq:deltaC}
\end{equation}
where ${\hat\omega}_{ss'}^{-1}(k)$ is the matrix inverse of 
${\hat\omega}_{ss'}(k)$.
This expression was used for the DRL theory also.
One problem though with Eq.(\ref{eq:deltaC}) for water is that the matrix
inverse of ${\hat\omega}_{ss'}(k)$ becomes singular as $k\rightarrow 0$. 
This singularity makes the algorithm not very stable. However, a simple
solution to this problem was found by cropping ${\hat\omega}_{ss'}^{-1}(k)$
near $k=0$. This cropping was done by setting the value of ${\hat\omega}_{ss'}(k)^{-1}$ for $k<k_{min}$ equal to its value at $k_{min}$, which is
proportional to
the inverse molecule size. For water, $k_{min}\simeq 0.5$ \AA$^{-1}$.

A new value for ${\hat c}_{es}(k)$ was then determined by mixing in a
fraction of $\Delta{\hat c}_{es}(k)$, this amount being 10-50\% typically.  
This whole procedure was then repeated - with one exception.
The one exception is
that the value of the charging integral in Eq.(\ref{eq:charging}) 
was held fixed until convergence was obtained on ${\hat c}_{es}(k)$.
At that time the charging integral was recomputed. A new value for
the charging integral was a mixture of its recomputed and old values, typically
at a ratio of 1:9. This second outer loop was then continued
until convergence on ${\hat c}_{es}(k)$ and thus $g_{es}(r)$ was obtained.
		
Since a variation of Eq.(\ref{eq:deltaC}) is also used for the RO-RPA
theory,\cite{donley05} this cropping of ${\hat\omega}_{ss'}^{-1}(k)$ was done for that theory too.

\bibliography{paper}
\end{document}